\documentclass[twocolumn]{aastex61}
\usepackage[T1]{fontenc}
\usepackage{apjfonts} 
\usepackage[figure,figure*]{hypcap}
\usepackage{systeme}
\usepackage[ruled,noline,noend,linesnumbered]{algorithm2e}
\usepackage{amsmath,graphicx}
\usepackage{listings}
\newcommand{\D}{{\sc digest2}}
\newcommand{\DD}{D_2}

\shorttitle{The \D NEO Classification Code}
\shortauthors{Vere\v{s} et al.}

\begin{document}

\def\thefootnote{*}\footnotetext{These authors contributed equally to this work}\def\thefootnote{\arabic{footnote}}
\title{Improvement of \D{} NEO Classification Code - utilizing the Astrometry Data Exchange Standard}

\author[0000-0002-5396-946X]{Peter~Vere\v{s} 1$^*$} 
\affiliation{Harvard-Smithsonian Center for Astrophysics, 60 Garden St., MS 15, Cambridge, MA 02138, USA}
\correspondingauthor{Peter~Vere\v{s}}
\email{peter.veres@cfa.harvard.edu}

\author[0000-0003-2586-2697]{Richard~Cloete 1$^*$}
\affiliation{Harvard-Smithsonian Center for Astrophysics, 60 Garden St., MS 15, Cambridge, MA 02138, USA}

\author[0000-0002-0439-9341]{Robert~Weryk}
\affiliation{Physics and Astronomy, The University of Western Ontario,1151 Richmond Street, London ON N6A 3K7, Canada}

\author{Abraham Loeb}
\affiliation{Harvard-Smithsonian Center for Astrophysics, 60 Garden St., MS 15, Cambridge, MA 02138, USA}

\author[0000-0001-5133-6303]{Matthew J. Payne}
\affiliation{Harvard-Smithsonian Center for Astrophysics, 60 Garden St., MS 15, Cambridge, MA 02138, USA}

\begin{abstract}

We describe enhancements to the $\D$ software, a short-arc orbit classifier for heliocentric orbits. \textit{Digest}2 is primarily used by the Near-Earth Object (NEO) community to flag newly discovered objects for a immediate follow-up and 
has been a part of NEO discovery process for more than 15 years. 
We have updated the solar system population model used to weight the $\D{}$ score according to the 2023 catalog of known solar system orbits and extended the list of mean
uncertainties for 140 observatory codes. 
Moreover, we have added Astrometry Data Exchange Standard (ADES) input format support to $\D{}$, which provides additional information for the astrometry, such as positional uncertainties for each detection. 
The $\D{}$ code was also extended to read the roving observer astrometric format as well as the ability to compute a new parameter from the provided astrometric uncertainties ($RMS'$) that can serve as an indicator of in-tracklet curvature when compared with tracklet's great-circle fit RMS. Comparison with the previous version of $\D{}$ confirmed the improvement in accuracy of NEO identification and found that using ADES XML input significantly reduces the computation time of the $\D{}$.

\end{abstract}


\section{Introduction}
\label{SECN:INTRO}
During the last two centuries, our ability to discover minor planets has changed significantly. Through visual telescopic observations in the nineteenth century, the discovery of photography and the utilization of photographic plates, to the introduction of charge-coupled devices late in the twentieth century, each technological step has rapidly increased the rate of discovery. In 2023, the number of discoverd minor planets reached 1.25 million, up from about 700000 in 2015.

In the last 25 years, the Solar system inventory surged due to dedicated Near-Earth Object (NEO) surveys that were undertaken after the 1998 United States Congressional mandate set the goal to discover $90\%$ of NEOs larger than 1~km (the Spaceguard Survey, \citet{Morrison92}), and the successive 2005 Congressional mandate to discover $90\%$ of NEOs larger than 140~m (George E. Brown, Jr. NEO Survey Act\footnote{Section 321 of the NASA Authorization Act of 2005 (Public Law No. 109-15)}). 
Modern NEO surveys rely on wide-field telescopes equiped with mosaic CCD cameras and large focal planes such as Pan-STARRS \citep{Kaiser02,Chambers16}, the Catalina Sky Survey \citep{Larson03}, and the Zwicky Transient Facility  \citep{Bellm19}, which strive to cover as much of the sky per night as possible, to as great a depth as possible. 
Surveys typically attempt to re-image the same area four times within a relatively small time window (e.g., one hour), allowing moving objects to be identified and their detections linked into a ``tracklet''. 
Detected moving targets are either attributed to known objects or reported to the Minor Planet Center\footnote{\url{https://minorplanetcenter.net/}} (MPC) as new object candidates. 

Typically, it is only by extending the short arc, which consists of a few positions, over several nights that an object can be designated and its orbit determined.

However, due to the limited availability of telescopic assets and the rapid rate at which new objects are discovered, it is not possible to immediately conduct follow-up observations on all of them.
For more than 15 years, the community has used a scoring tool $\D{}$\footnote{\url{https://bitbucket.org/mpcdev/digest2/overview}} to classify unknown tracklets and post prospective NEO candidates to the MPC's Near-Earth Object Confirmation Page (NEOCP\footnote{\url{http://minorplanetcenter.net/iau/NEO/toconfirm_tabular.html}}). 
Given an input tracklet and a selected orbit category, the code computes a numeric score ranging from 0 to 100. 

The score is calculated by counting the number of variant orbits that represent the input tracklet. The counts are down-weighted by the expected number of orbits and undiscovered orbits of a given size in any given population bin. 
Although the software allows users to select their own orbit category, it is worth noting that the majority of users utilize $\D{}$ specifically for calculating NEO scores: tracklet scoring above 65 is eligible to be poseted on NEOCP.
Detailed work on $\D{}$ by \citet{Veres19} describes the concept and functionality of the scoring mechanism and also outlines its current caveats, disadvantages and future developments. In this work, we present an updated \D{} utilizing the modern Astrometry Data Exchange Standard - ADES \citep{Chesley17}, an updated population model, an updated list of assumed astrometric uncertainties, and an intra-tracklet residual analysis.

\section{$\D{}$ improvements}
\label{s:Improvements}

\subsection{Updated population model}
\label{ss:Improvements:pop}

The $\D{}$ code requires a population model in the form of binned data that is used to weight the variant orbits produced for a given tracklet \citep{Veres19}. There are two populations: The full population model and the undiscovered population model. 

The full population model represents the complete synthetic solar system based on the Pan-STARRS Synthetic Solar System Model (S3M) of \citet{Grav11}, consisting of
over 14 million simulated Keplerian orbits represented by perihelion distance ($q$), eccentricity ($e$), inclincation ($i$) and absolute magnitude ($H$) that roughly represent the size of the object. The full population model yields the `'Raw' $\D{}$ score. 

The undiscovered population model is derived by subtracting the current catalog of discovered orbits from the S3M model in four dimensional bins (q,e,i,H), yielding a `NOID' $\D{}$ score. 
However, the population model distributed with $\D{}$ code is updated infrequently. Decreasing the population of undiscovered objects in a given orbital element-size bin would increase the score for orbits in other bins, likely for the smaller objects, expecting the larger objects to be completed first.

Before the publication of \citet{Veres19}, the model was last updated in 2015 when the known catalog consisted of 700000 orbits.  The subsequent update of the population model came in 2021 when the catalog listed 1.07 million objects. In this work, we update the model based on the MPC orbit catalog (MPCORB\footnote{\url{https://www.minorplanetcenter.net/iau/MPCORB.html}}) from 2023-02-26, which contains 1.26 million orbits.

\subsection{Astrometric uncertainties}
\label{ss:Improvements:unc}
For a given tracklet, variant orbits are constructed from the synthetized tracklet end-points that are dithered based on the astrometric uncertainty \citep[for further details, see][]{Veres19}. 
The uncertainty value is either assumed based on the observatory code\footnote{\url{https://minorplanetcenter.net//iau/lists/ObsCodesF.html}}, or a default value of 1.0 arc-second is used. Historically, only a relatively short list of uncertainties were provided: the 2015 version of $\D{}$ was distributed with a list of 17 observatory code uncertainties that were derived from the mean uncertainty based on the MPC orbit fitting of all known orbits with reported astrometry submitted from a given observatory code. The 2021 version contained 35 observatory code with updated uncertainty values.

In response to the escalating volume of incoming astrometry and the integration of new observatory codes, we have compiled a more extensive list of 140 observatory codes, drawn from major asteroid surveys and prominent NEOCP reporters, in this study. To establish a new population model, we calculated the mean uncertainties using perturbed orbit solutions derived by $OrbFit$\footnote{\url{http://adams.dm.unipi.it/~orbmaint/orbfit/}}, a system that has superseded the in-house MPC orbit fitting methods. These uncertainty values are portrayed as mean figures in the magnitude bins with the highest number of submissions, thereby offering the most conservative estimate for faint detections. Consequently, our expanded list now includes uncertainties for each of the 140 observatory codes.

\subsection{ADES and the astrometric uncertainty}
\label{s:Improvements:ADES}
For about three decades, astrometric observations of minor planets and comets have been submitted to MPC and then distributed and published using the so-called MPC1992\footnote{\url{https://minorplanetcenter.net/iau/info/OpticalObs.html}} 80-column ASCII format.

The concentrated format encapsulates information on a given designation, observation time (epoch), right ascension, declination, magnitude, band, astrometric reduction catalog, position of the observer in a form of an observatory code and additional information on observing mode (e.g. CCD), and observing notes or a program code for each astrometric position on a single 80-character line (see Appendix~\ref{app:example_tracklet}). Astrometry submitted from space-based observatories, by radar, or by observers without permanent topocentric location (roving observers), provide additional information on a second 80-column line, describing the geocentric location or radar-specific data.

In 2015, the International Astronomical Union adopted the Astrometry Data Exchange Standard (ADES) as a new format for the distribution and submission of astrometric data. This format, which uses eXtensible Markup Language (XML) instead of the limited 80-character format, allows for significantly more information to be included with each astrometric data point. Notably, ADES includes additional fields that improve the accuracy of astrometry, such as reported astrometric uncertainties (see Appendix~\ref{app:example_tracklet}).
Prior to the recent update of $\D{}$, the only acceptable input format for astrometry data was MPC1992.
In recent years, submission in ADES format has become dominant (Table ~\ref{t:MPC-ades_submissions}), and the MPC is continuously working on the distribution of published ADES-formatted astrometry: currently MPEC-published astrometry and mid-month batches are also distributed in the ADES format, and fields like uncertainties are available in the MPC observations database replicated at Small Bodies Node of NASA's Planetary Data System\footnote{\url{https://sbnmpc.astro.umd.edu/MPC_database/statusDB.shtml}}. 

In this work we have added functionality that allows $\D{}$ to read XML (any input file name with the extension .xml) and ingest the reported astrometric uncertainties. 
The previous version of $\D{}$ retrieved the uncertainty from an input configuration file containing a list of 35 observatory codes with a mean uncertainty for each code, or assumed the default astrometric uncertainty of 1 arc-second if the observatory code of the tracklet was not in the configuration file. Even though the input file values were derived from statistical analysis of historical data of the most prolific surveys and observatory codes, in reality the conditions for each observation differ significantly: astrometric uncertainty decreases as a function of increasing brightness, and the signal-to-noise ratio of the same detection can change from exposure to exposure if the observing conditions change. Thus, using reported and measured uncertainties greatly improves the generation of variant orbits and thus providing the $\DD{}$.

XML-provided uncertainties submitted by an observer could contain unreasonably high or low values. As such, we have implemented a floor and ceiling for reported astrometric uncertainties for obscodes that are in the configuration file with an expected astrometric uncertainty: the XML-provided uncertainties must be within 0.7 and 5-times the tabular (expected) value. The value of 0.7 was selected by analysing residuals of 10 most active surveys and for their bright astrometry: astrometric uncertainty decreases as a function of signal-to-noise and our analysis showed that for the bright end, the RMS is about 0.7-times smaller with respect to the faint end. The 5-times limit represents the 5-sigma threshold. In addition to the limits, the uncertainty is selected as a maximum of the reported or configuration astrometric uncertainty and the RMS of the great-circle-fit, similarly to the previous functionality of the code for the MPC1992 input.

\begin{table}
\small
\begin{center}
\caption{Fraction of ADES-submitted astrometry as a function of observation time and the total number of astrometric positions published by MPC or present in the isolated tracklet file.}
\tabcolsep=0.11cm
\begin{tabular}{c|c|c}
\hline
year &  fraction of ADES &  Observations\\
\hline
2018 &  13\%  &23.9 million \\
2019 &  40\%  &33 million \\
2020 &  48\%  &42.6 million \\
2021 &  82\%  &32 million \\
2022 &  91.5\%& 34 million \\
\hline
\end{tabular}
\label{t:MPC-ades_submissions}
\end{center}
\end{table}

Our XML implementation reads reported astrometric uncertainties and uses the first and last reported positions for the generated end-points and all reported values for work described in \S\ref{s:Data}. If the uncertainties are not provided, $\D{}$ will use either a default value read from the configuration file for the observatory code, or the default 1.0 arc-second. The code will parse the XML format only for input files with the `.xml' suffix. 

If the input file has any other suffix or no suffix at all, the code expects the MPC1992 format.

\subsection{Roving observer}
Observers who do not have a permanent topocentric position defined by an observatory code can still submit astrometry to MPC in the so-called roving observer format\footnote{\url{https://minorplanetcenter.net/iau/info/RovingObs.html}}. The MPC1992 roving observer provides the astrometry but instead of the observatory code defining the topocentric location, the geodetic coordinates are submitted directly on the second line of the MPC1992 ASCII. One of two observatory codes must be used for the roving observer format: 247 (roving) or 270 - the roving format utilized by the Unistellar network \citep{Marchis20}. Roving observers were not supported in the previous versions of $\D{}$. In this work, we added support for roving observers as well as for ADES astrometry.

\section{Data validation and code verification}
\label{s:Data}

We selected four independent data sources to test the performance of the updated population model, configuration file and the ADES-formatted astrometry (Table ~\ref{t:data-counts}). To test the true and false-positive rate of identifying NEOs and main-belt asteroids (MBAs), the first two data sets consisted of known NEO and MBA tracklets observed between `2020-01-01' and `2023-01-01' with known orbits, each having at least three detections and fainter than $V=19.5$ magnitude. The selection of the magnitude threshold was motivated by the fact that the background main-belt population is nearly complete at the specified apparent magnitude. Therefore, all bright objects are of interest, as they could either be NEOs, comets, distant objects, or other atypical objects that are easily distinguished as noteworthy\footnote{The MPC permits the submission of bright objects to NEOCP without considering their $\D{}$ score.}.
Also, an absolute magnitude ($H$) threshold was set for the orbits, assuming the large objects are almost complete at a given $H$: $H>20$ for NEOs and $H>13$ for non-NEOs.

The third data set represents the Isolated Tracklet File (ITF) of unidentified objects observed between `2020-01-01' and `2023-01-01', having at least three detections, and without any magnitude limit.

The last data set consists of NEOCP discovery (initial) tracklets observed between `2019-02-26' and `2023-02-19', regardless of their magnitude or final attribution. Roughly $55\%$ belonged to NEOs, $30\%$ to non-NEOs and the remaining $15\%$ to either unidentified tracklets ending up in the ITF, being identified as artificial or deleted by the observer. 

All data were submitted with their astrometric uncertainties in the ADES format.
Thus, we were able to perform the $\DD{}$ calculation on both the ADES-XML formatted data with the reported uncertainties, \emph{and} on version of the data converted to MPC1992 format (and hence lacking astrometric uncertainties).
The NEOCP sample, however, also contained data without ADES submission - only about 15000 out of 22000 NEOCP tracklets were submitted in the ADES XML format.  Nevertheless, we were motivated to explore the full NEOCP sample.

The four data sources were diverse and we expected that the NEO $\D$ score would vary for each. 
For instance, we would expect known NEOs to have large $\DD$, while most known MBAs would yield very low scores.

The NEOCP sample is a mixture of NEOs and non-NEOs, but all of them, except for a small sample of comets, were posted to the NEOCP because they had $\DD>65$. 
Thus, the data set is ideal for testing the MPC-adapted thresholds where $\DD>65$. 
The ITF data represent objects having unknown orbits and most are likely part of the faint background MBA population, thus their $\DD{}$ is expected to be low.

\begin{table}
\small
\begin{center}
\caption{Number of tracklets, detections and time-range of the tested astrometry data sets. Only about 70\% of NEOCP data had ADES information.}
\tabcolsep=0.11cm
\begin{tabular}{c|c|c|c}
\hline
Type & Tracklets  & Observations & Date-range \\
\hline
known NEOs & 30829&110989 & 2020-01-01 2023-01-01\\
known MBAs &873239 & 3430460& 2020-01-01 2023-01-01 \\
ITF &90702&  346112& 2020-01-01 2023-01-01\\
NEOCP & 22100*& 83125* & 2019-02-26 2023-02-19\\
\hline
\end{tabular}
\label{t:data-counts}
\end{center}
\end{table}

\subsection{Updated population model and astrometric uncertainties}

\begin{figure}%
\centering
 \includegraphics[scale=0.5]{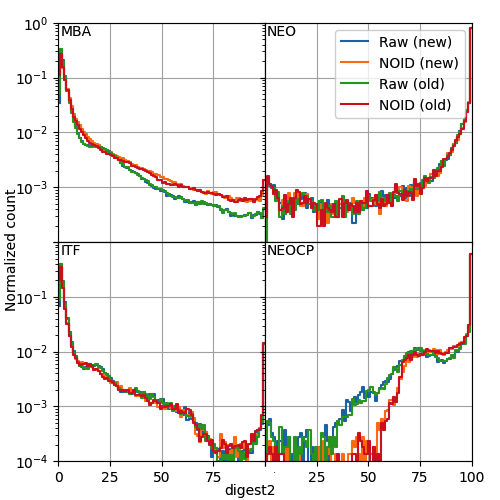}
    \caption{Histograms of $\D{}$ of `new' and `old' values for Raw and NOID NEO $\DD{}$ when using the previous and new population model and configuration file, executed on four data sets: known NEO, known MBA, ITF and NEOCP and on MPC1992 astrometry data format.}
    \label{fig:histograms_d2_pop}
\end{figure}

We computed $\D{}$ scores for the four data sets defined in \S\ref{s:Data} using the old population model and configuration file, as well as with the new population model (see \S\ref{ss:Improvements:pop}) and the new configuration file (see \S\ref{ss:Improvements:unc}). In all cases, we used the $repeatable$ keyword, ensuring the randomness factor was disabled, thus preventing different $\D$ from generating different scores for the same tracklet. Table~\ref{t:hist_pop} and Figure~\ref{fig:histograms_d2_pop} show the resulting Raw and NOID NEO $\D{}$ scores. The histograms clearly show the difference between the data sources: $\D{}$ distinguishes NEOs and MBAs almost as a binary classifier. The $\DD$ of nearly all NEO tracklets are above $\DD{}>65$ while the MBAs are below. The ITF data sample shows $\D{}$ behavior similar to the MBAs, thus showing the ITF tracklets most likely all belong to unknown MBAs. NEOCP tracklets are also almost all above the $\DD{}>65$ but this is expected: the threshold is a requirement for posting to NEOCP. There is a small fraction that had $\DD{}<=65$ but those mostly represent new comet discoveries that are allowed on NEOCP regardless of their $\D{}$, or are prediscovery tracklets that were added to NEOCP as an object from the ITF. The histograms (Figure~\ref{fig:histograms_d2_pop}) show no fundanental difference between the `old' and `new' set of population model and configuration file when comparing Raw and NOID NEO $\DD{}$ scores. We focused on NOID $\D{}$ primarily because the NOID is used as a NEOCP threshold score.

\begin{table}
\small
\begin{center}
\caption{Percentage of tracklets above or below a given NOID $\D{}$ thresholds and a tested population for a new population model and a new configuration file.}
\tabcolsep=0.11cm
\begin{tabular}{c|c|c|c}
\hline
Type & <=65 & >65 & =100 \\
\hline
NEO & 3.7 &96.3 &74 \\
MBA & 98 & 2 & 0.0005  \\
ITF &  98& 2 & 1.3\\
NEOCP &3& 97 & 57\\
\hline
\end{tabular}
\label{t:hist_pop}
\end{center}
\end{table}

\begin{figure}%
   \includegraphics[scale=0.5]{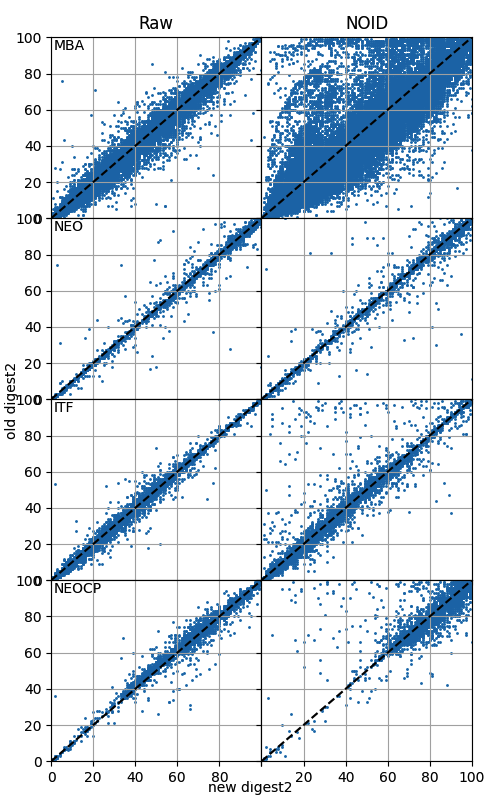}

    \caption{Scatter plots of `new' and `old' values for Raw and NOID NEO $\DD{}$ when using the previous and new population model and configuration file, executed on four data sets: known NEO, known MBA, ITF and NEOCP and on MPC1992 astrometry data format.}
    \label{fig:population_scatter}%
\end{figure}

\begin{figure}%
\centering
\includegraphics[scale=0.5]{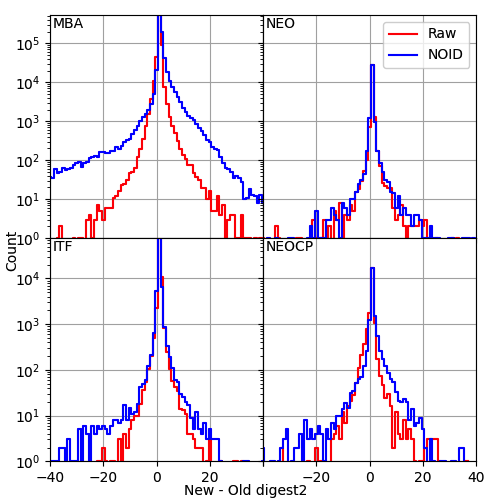}

\caption{Difference between the `new' and `old' values for Raw and NOID NEO $\DD{}$ the previous and new population model and configuration file, executed on four data sets: known NEO, known MBA, ITF and NEOCP and on MPC1992 astrometry data format.}
    \label{fig:population_histogram}
\end{figure}

Subsequently, we analyzed the difference between the `new' and `old' $\D{}$ in detail. 
We anticipated observing diverse $\DD{}$ values, with a greater variance in the NOID score for the `new' model. 
The Raw value relies on the complete synthetic Solar System model, which remains unchanged between the `old' and `new' approaches. Therefore, in the Raw comparison, only the expanded and revised list of uncertainties impact the resultant $\D{}$ scores.
In the case of NOID, the resulting effect is a convolution of the updated population model, because NOID reflects the score with respect to the undicovered portion of the Solar System population plus the updated uncertainties in the configuration file. The scatter plots in Figure~\ref{fig:population_scatter} display a wider scatter for the NOID $\D{}$ score. There is almost no difference between Raw and NOID $\Delta\DD{}=old-new$ for the MBA, NEO and ITF data samples for $\Delta\DD{}$ (Figure~\ref{fig:population_histogram}) and most of the  tracklets did not have any change in the score: the peak difference is always at zero. The $\Delta\DD{}$ is symetrical for all data samples.
The NOID differs from Raw mostly for MBAs, implying that the population update was mostly affecting MBAs objects, and the effect of the population update was more prominent than the per-obscode uncertainties.

Table~\ref{t:pop_compare_NOID} shows the fraction of tracklets that had no change in their NOID $\D{}$ score, the fraction of tracklets that experienced a small change (up to $\Delta\DD{}<=5$), as well as on how many tracklets were downgraded to below ($\downarrow$) or upgraded to above ($\uparrow$) the threshold of $\DD{}=65$. 

A marginally greater number of Near-Earth Objects (NEOs) were promoted above the threshold compared to those demoted below it. Additionally, slightly more Main-Belt Asteroids (MBAs) were downgraded than upgraded. This pattern suggests that the $\D{}$ score is enhancing the accuracy of true-positive NEO classification.
However, this pattern does not hold universally. Twice as many known ITF tracklets had their $\DD{}$ score increased above the threshold compared to those decreased. In contrast, for NEOCP candidates, twice the number had their score downgraded.

An in-depth analysis of the downgraded and upgraded NEOCP tracklets revealed the following: 111 tracklets were downgraded below 65 and consequently would not appear on the NEOCP, while 45 had their $\D{}$ score raised above 65. These findings could suggest a diminished performance of the enhanced population model and the uncertainty model.
However, upon closer inspection, none of the 111 downgraded tracklets were NEOs. Of them, three were identified as numbered comets, 12 remained as ITF tracklets, 16 were removed by the submitter, and the remaining were known non-NEOs.

Therefore, not a single NEO was downgraded. One might contend that the ITF tracklets could be unknown NEOs, however, all of the aforementioned ITF tracklets had only intermediate NEO $\D{}$ scores - in the high sixties and seventies. Therefore, the probability of these tracklets being NEOs was minimal (see Figure 1). Most of the ITF tracklets from the NEOCP in this sample were sourced from Pan-STARRS and upon further examination of the images, it was found that the tracklets were predominantly false-positives.

From the 41 upgraded NEOCP tracklets, only two remain in the ITF, two were deleted, one was identified as an NEO, and the rest were non-NEOs. This demonstrates that for the NEOCP sample, the enhancement would reduce the number of false-positive NEOCP candidates by approximately 50, while promoting one true NEO above the threshold.

\begin{table}
\small
\begin{center}
\caption{Analysis of the difference of the NOID $\D$ score, giving a fraction of the data being within the $\Delta$ of 0,1,3,5. Arrows show how many tracklets were  upgraded above 65 and downgraded below 65 when the new population model and configuration file were used with respect to the old settings.}
\tabcolsep=0.11cm
\begin{tabular}{c|c|c|c|c||c|c|c}
\hline
Type & 0 & <=1 & <=3 & <=5 & $\uparrow$ & $\downarrow$ & Total \\
\hline
known NEOs&90.3 & 98.4 & 99.3 & 99.6 & 36 & 30 &30823\\
known MBA&61.2 & 88.2 & 95.0 & 97.1 & 2014 &2241 &873209\\
ITF&83.5 & 98.0 & 99.3 & 99.7 & 105& 50 &90703\\
NEOCP&76.9 & 93.2 & 96.8 & 98.2 & 45 & 111 & 21765 \\
\hline
\end{tabular}
\label{t:pop_compare_NOID}
\end{center}
\end{table}

\subsection{Comparing ADES and MPC1992 data}

The $\D{}$ code has been updated to read XML, enabling support for floating point values (and thus increased precision), while also providing the data in a far more human-readable format than the current 80-column MPC1992 format. Furthermore, XML support brought with it a significant improvement in terms of data processing performance.
In contrast to the previous version of the code that required multiple parsing and arithmetic operations for each line, the XML input requires fewer operations and less memory manipulation.
This has resulted in a several-fold decrease in processing times for $\D{}$ (see Table~\ref{t:runtime}). Our experimentation was carried a ThinkMate server with a 16-core CPU \@3.9 GHz and 128 GB of RAM. Importantly, the $\D{}$ code supports multi-core processing, allowing it to scale with the number of available CPU cores.

\begin{table}
\small
\begin{center}
\caption{Runtime in minutes of $\D$ on the data sets from Table ~\ref{t:data-counts} on MPC1992 and ADES XML data format. $\D$ is significantly faster when running on XML input.}
\tabcolsep=0.11cm
\begin{tabular}{c|c|c|c}
\hline
Type &  MPC1992 & ADES XML & Improvement factor\\
\hline
NEOs & 3.3 & 0.6 & 5.4 \\
MBAs & 251.3 & 39.5 & 6.3\\
ITF  & 26.7&4.2 & 6.3 \\
NEOCP   &1.9 & 0.3  & 6\\
\hline
\end{tabular}
\label{t:runtime}
\end{center}
\end{table}

In the following comparison we executed the $\D$ program on the same data where one was provided in the MPC1992 format and the other in ADES XML format. In both cases we used the updated population model (\S\ref{ss:Improvements:pop}) and the configuration file with updated per-obscode uncertainties (\S\ref{ss:Improvements:unc}). The primary distinction between the two formats resides in the astrometric uncertainties. The ADES format offers these uncertainties directly. In contrast, the MPC1992 format provides a qualified estimate reliant on the configuration file and contingent upon the observatory code.

\begin{figure}%
\centering
\includegraphics[scale=0.5]{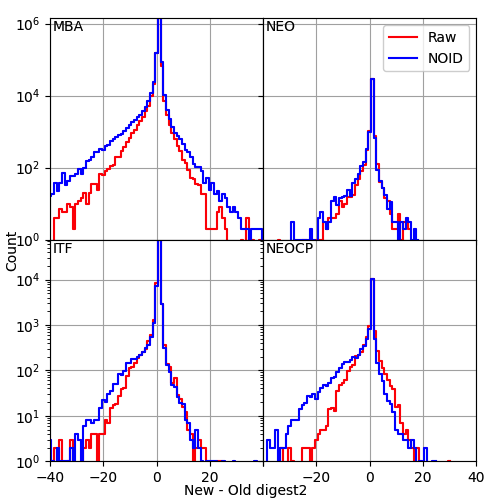}
\caption{Difference between the `new' and `old' values for Raw and NOID NEO $\DD{}$ comparing the MPC1992 with new population model and configuration file (old) and the ADES XML data, executed on four data sets: known NEO, known MBA, ITF and NEOCP and on MPC1992 astrometry data format.}
    \label{fig:population_histogram_xml}
\end{figure}

Table~\ref{t:pop_compare_NOID_xml} shows the difference between the ADES and MPC1992 NOID NEO $\D{}$ and how many tracklets were upgraded or downgraded with respect to the $\DD{}=65$ threshold. In comparison with Table~\ref{t:pop_compare_NOID} and Figure~\ref{fig:population_histogram_xml}, it is clear that the differences between MPC1992 and ADES results are smaller than between old and new population model and the configuration file comparison for MPC1992 input only. On the other hand it seems that many more tracklets are being downgraded than upgraded. 
This decreasing false-positive rate is potentially a good sign for MBAs and ITFs, but raises concerns for NEOs. The change is quite negligible for known NEOs (15 upgraded versus 55 downgraded out of more than 30000 tracklets) but the increase is one order of magnitude for the NEOCP tracklets. As mentioned earlier, NEOCP tracklets are a mixture of NEOs and non-NEOs and the community would strongly benefit from an improvement that would remove non-NEOs from the NEOCP. A closer look at 623 downgraded tracklets shows that only 22 ended up in the ITF and the remaining were designated. From those, only 16 were NEOs while vast majority were non-NEOs. We don't know the true orbits of the 20 ITF tracklets, but their NEO $\D{}$ were moderately high, therefore we expect these are unlikely to be NEOs. Thus, the ADES astrometry implementation in $\D{}$ greatly decreases the false-positive rate on the NEOCP.

\begin{table}
\small
\begin{center}
\caption{Analysis of the difference in the NOID $\D$ score when calculated using MPC1992 and ADES astrometry, giving a fraction of the data being within the $\Delta$ of 0,1,3,5. Arrows show how many tracklets were  upgraded above 65 and downgraded below 65 when comparing $\D{}$ score for MPC1992 and ADES astrometry format.}
\tabcolsep=0.11cm
\begin{tabular}{c|c|c|c|c||c|c|c}
\hline
Type & 0 & <=1 & <=3 & <=5 & $\uparrow$ & $\downarrow$ & Total \\
\hline
NEOs&91 & 96.6 & 99.3 & 99.7 & 15 & 55 &30823\\
MBAs&81.1 & 98.7 & 99.5 & 99.7 & 583 & 4105 &873209\\
ITF&83.0 & 99.0 & 99.5 & 100 & 28 & 250 &90703\\
NEOCP&68.2 & 97.4 & 98.7 & 98.8 & 41 & 623 & 15234 \\
\end{tabular}
\label{t:pop_compare_NOID_xml}
\end{center}
\end{table}

\subsubsection{ADES digest2 with input-forced astrometric uncertainties}

In this version of the code, we have added an option to execute the $\D$ code with user-provided uncertainties and without any threshold. 
This addition empowers submitters to calculate the score without any restrictions by employing the keyword $noThreshold$. 
We have evaluated the behavior of the $\D$ score using the same four data samples (NEO, MBA, ITF, NEOCP).

\begin{figure}%
\centering
\includegraphics[scale=0.5]{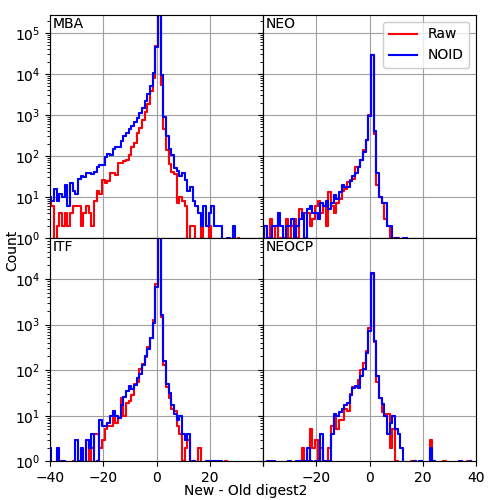}
\caption{Difference between the `new' and `old' values for Raw and NOID NEO $\DD{}$ comparing the MPC1992 with new population model and configuration file (old) and the ADES XML data, executed on four data sets: known NEO, known MBA, ITF and NEOCP and on MPC1992 astrometry data format, but with the noThreshold keyword, using reported XML astrometric uncertainties as reported.}
    \label{fig:population_histogram_xml1}
\end{figure}

Figure~\ref{fig:population_histogram_xml1} illustrates the variance between the `new' (e.g., $noThreshold$) and `old' $\DD{}$ for both Raw and NOID values. An immediate comparison of Figure~\ref{fig:population_histogram_xml1} with Figures \ref{fig:population_histogram} and \ref{fig:population_histogram_xml} underscores a significant alteration in behavior: the distribution no longer displays symmetry. Rather, the distribution is heavily skewed towards negative values, indicating that the $noThreshold$ score is consistently lower.

Our hypothesis attributes this to an underestimation of the reported uncertainties, leading to diminished resultant scores. 
Such behavior implies that the $noThreshold$ mode should \emph{not} serve as the primary tool for large-scale data analysis and real-time orbit-type classification. Instead, it should be regarded as an experimental instrument for research purposes. 

\section{Finding low \D{} NEOs by in-tracklet curvature}
\label{s:curvature}
One of the conclusions of \cite{Veres19} was that the $\DD{}$ of certain NEOs could be low. It is expected that if a NEO is relatively far from Earth (the observer) and near its aphelion in the Main Belt, then its apparent rate of motion would be slow and could mimic the motion of main belt objects, particularly when observered near opposition. On the other hand, when a NEO is near the observer, its apparent rate of motion is faster than that of any Main Belt object and thus, its $\DD{}$ is high. However, there are cases when the NEO is both close and has a slow rate of motion.
This happens for objects on Earth-like orbits, objects approaching the Earth at a particular geometry or some objects on direct impact trajectory \citep{Veres09}. One prominent example was $2019$\,OK \citep{Wainscoat22}, where on 2019-07-25, a relatively large NEO ($H=23.3$) came within $100000\, km$ of the Earth. The NEO was discovered only a day before the event, as a very bright unknown object ($V=14$) with a $\DD{}=100$ causing it to be posted to the NEOCP. Surprisingly, a subsequent search of the image archives revealed that the object had been observed on multiple occasions, stretching back to a month prior, by both Pan-STARRS and ATLAS \citep{Tonry18}. However, each instance exhibited a rate of motion that was too slow to be identified by their automated detection pipelines.

The basic operational principle of $\D{}$ is the use of a pair of end-points that are used for the construction of the velocity vector and variant orbits matching the observed tracklet, thus generating the $\DD{}$ score. However, if the object is close to the observer, the mutual geometry changes rapidly and thus, the effect can be observed even within the tracklet - the motion diverges from a great circle motion. If the number of detections per tracklet is three or more, such an effect can be significant and visible when compared with its great-circle fit. $\D{}$ computes the residual RMS of great-circle fit of each tracklet according to equation \ref{eq:RMS}:

\begin{equation}
\label{eq:RMS}
RMS=\sqrt{\frac{\sum{\Delta\alpha_{i}^2+\Delta\delta_{i}^2}}{N}}
\end{equation}

\begin{figure}[hpt]
\centering
  \includegraphics[width=5cm]{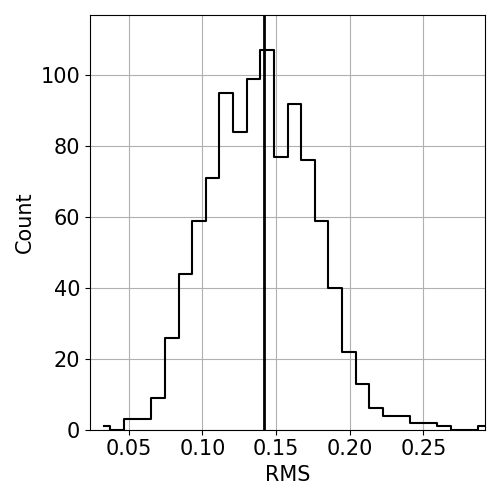}
  \caption{Simulation of a 4-detections tracklet: each position was generated as a random number with a mean on a great-circle and standard deviation of $\sigma=0.1\arcsec$, in 1000 trials. For each trial the $RMS$ is computed and the resulting histogram is displayed. The $RMS1=0.14\arcsec$ is derived from the expected uncertainty of $0.1\arcsec$}
    \label{fig:rms_trials}
\end{figure}

Previously, without knowledge of the true astrometric uncertainty, one could only make an assumption, making it challenging to distinguish measurement uncertainties from the in-tracklet parallax effect. However, with the introduction of reported astrometric uncertainties in the ADES astrometry, we can compare the Root Mean Square (RMS) of the great-circle fit with the RMS of the reported uncertainties using the same equation (\ref{eq:RMS}), here denoted as $RMS'$. If the RMS of the fit is statistically and significantly larger than the expected great-circle fit $RMS'$ based on the reported uncertainties, then the motion would suggest a small geocentric distance. This finding could serve as a potential marker for low-digest NEOs that might be posted to the NEOCP earlier. As such, we implemented this functionality such that $RMS'$ is returned by $\D{}$ when the keyword $rmsPrime$ is provided in the configuration file. The $RMS'$ is computed from XML-provided uncertainties. If the value is not provided, it is substituted with the observatory-specific assumed uncertainty from the configuration file (refer to \S\ref{ss:Improvements:pop}) or it defaults to a value of $1.0\arcsec$. 
Figure \ref{fig:rms_trials} illustrates our simulation results for the great-circle fit RMS of four detections with known measured uncertainties. These uncertainties are randomly distributed according to a normal distribution, where the $1-\sigma$ is represented by uncertainties in both the $\alpha$ and $\delta$ directions.
The figure suggests that within 3-sigma probability the $RMS$ < 2$RMS'$.

We performed an evaluation of the true-positive and false-positive rates of detections with large RMS values for ADES-submitted astrometry on the data sets detailed in Table~\ref{t:data-counts} and with the $noThreshold$ keyword. The table below shows the number of tracklets with large RMS ($>2RMS'$) values having $\DD<65$, which were computed using the ADES XML astrometry with the updated population model and uncertainty file.
Table~\ref{t:curvature_rms} summarizes our search for low-digest (NEO NOID) $RMS$>2$RMS'$ candidates, and indicates that there were only a handful for known NEOS (eleven) found; two example tracklets are shown in Appendix~\ref{app:example_curvature}.

\begin{table}
\small
\begin{center}
\caption{Positive curvature detection based on GCR RMS and provided astrometric uncertainties. The number of potential candidates is very small.}
\tabcolsep=0.11cm
\begin{tabular}{c|c|c|c}
\hline
Type & Tracklets & Fraction of $\DD{}<65$& RMS detection\\
\hline
NEOs & 30828 & 3\% & 11 \\
MBAs &873275 & 97\% & 1984 \\
ITF &97693 & 98\% & 271 \\
NEOCP & 15235 & 6\% & 11 \\
\hline
\end{tabular}
\label{t:curvature_rms}
\end{center}
\end{table}

For MBAs, there were almost 2000 tracklets which seems large, but small when compared to the total number of MBAs. 
We consider this MBA detected curvature to be a false-positive: MBAs are too far away to have meaningful curvature in a short arc. 

Out of those 2000 tracklets, over half (more than 1000) had magnitude information missing for at least one detection within their tracklet. 
This absence is often a signal of issues encountered during the astrometric solution, suggesting a potential problem in the astrometry not reflected in the reported uncertainties. An additional indicator of the affected astromeric quality could be the publishable note\footnote{https://minorplanetcenter.net/iau/info/ObsNote.html}: we found dozens of observations with the note `image tracked on object motion', almost one hundred with `stacked' and two with `hand measurement of CCD image'.
Of the remaining thousand tracklets, we analyzed how the reported uncertainty compared to the assumed mean uncertainty for a given observatory code from the configuration file. This analysis focused on the most frequently utilized observatories: I41, G96 V00, Pan-STARRS (F51, F52) and ATLAS (T05, T08, W68). 
Table~\ref{t:MBA_false_positive_curvature} reveals that most of the tracklets exhibiting false-positive curvature had reported astrometric uncertainties much lower than expected and were thus most likely underestimated.

\begin{table}
\small
\begin{center}
\caption{Analysis of false-positive curvature from the MBA sample. The table shows the mean magnitude, $RMS'$ computed from the per-obscode configuration file instead of the reported XML astrometric uncertainty, and the fraction of $RMS'$ derived from reported XML with respect to the mean expected value.}
\begin{tabular}{c|c|c|c}
\hline
obscode & $\overline{mag}$ & $\overline{\sigma}$ & $Fraction<\overline{\sigma}$ \\
\hline
I41 & 20.1 & $0.31\arcsec$ & 95\% \\
G96 & 20.2 & $0.41\arcsec$ & 98\% \\
V00 & 21.1 & $0.24\arcsec$ & 93\% \\
ATLAS & 19.6 & $0.56\arcsec$ & 100\% \\  
PS  & 21.5 &  $0.18\arcsec$ & 95\% \\
\hline
\end{tabular}
\label{t:MBA_false_positive_curvature}
\end{center}
\end{table}

Next we looked into the 271 large RMS low-digest ITF tracklets. Out of these, 50 belonged to ATLAS (T05, T08, M22, W68) and were primarily bright, with some registering at $V=11$. An examination of their FITS images revealed that the majority were false detections, rather than actual objects. This highlights a known issue: a small but undefined fraction of ITF detections are artifacts or glitches, rather than real objects. This occurs because the majority of automatically processed and submitted astrometry from large surveys does not undergo human vetting, allowing some false detections to slip through the detection software of the submitters. A small number were known objects with very large errors on a single detection, preventing attribution to a known object. Another subset comprising 89 tracklets were missing magnitude information for some of their detections, suggesting a problem with their astrometry. 

After conducting a manual investigation and assessing ATLAS images, we were left with approximately half of the potential large RMS tracklets from the ITF.

Of all tracklets within the NEOCP, only eleven showed positive curvature: three were comets, introducing bias to the astrometry due to coma or cometary activity; three remain in the ITF and potentially represent real objects given their positive curvature; one was an artificial satellite likely having real curvature due to it being close to Earth; one was a NEO and three were non-NEOs.

This demonstrates that the comparison of the GCR RMS with the reported uncertainties is a promising method for identifying low-digest NEOs, but this is contingent on the accurate reporting of uncertainties and unaffected astrometry. Furthermore, in cases of potential curvature detection, we strongly advise astrometry submitters to visually examine their images. This is because many of the submitted and scrutinized potential tracklets were found to be false. Our suspicion on understimated reported astrometric uncertainty recommends not using the $noThreshold$ keyword in regular $\D{}$ operations and to rely on the default configuration that uses floor and ceiling for the ADES-submitted uncertainties.

\subsection{Along and cross-track acceleration in a tracklet}

Without attempting to fit a preliminary orbit of a tracklet, we can approximate its motion as an initial position and skyplane speed towards a given position angle, which then accelerates in both the along and cross-track directions. We may expect that the vast majority of tracklets do not show any meaningful acceleration during their very short arcs which are typically minutes up to one hour. However, if an object is very close to the Earth, its acceleration can be significant.

With three or more detections (i.e., a system of $2N$ non-linear equations, with $N$ being the number of detections), we fit six variables (Equation~\ref{eg:system}) using a Levenberg-Marquardt Method: initial right ascension ($\alpha_{0}$) and declination ($\delta_{0}$), position angle ($A$), initial apparent rate of motion ($\bar{v}$), along-track ($\bar{a_{\parallel}}$) and cross-track ($\bar{a_{\perp}}$) accelerations, with the fit positions given by $\alpha_{i}$,$\delta_{i}$ at relative times of observation $T_{i}$.

\begin{equation}
    \begin{aligned}
        \alpha_{i} &= \alpha_{0} + T_{i}  \left[\left(\bar{v} + T_{i}  \frac{\bar{a_{\parallel}}}{2} \right) \sin(A) - \left(T_{i} \frac{\bar{a_{\perp}}}{2}\right)  \cos(A) \right]/\cos(\delta_{0}), \\
        \delta_{i} &= \delta_{0} + T_{i} \left[ \left(\bar{v} + T_{i}  \frac{\bar{a_{\parallel}}}{2} \right) \cdot \cos(A) + \left(T_{i} \frac{\bar{a_{\perp}}}{2}\right)  \sin(A) \right].
    \end{aligned}
    \label{eg:system}
\end{equation}

Each measurement is weighted by assumed uncertainties based on \citep{Veres17a}, since the vast majority did not have ADES-provided uncertainties. Using the covariance matrix of each fit, we can estimate how statistically significant each variable is. Because the along-track error is likely larger than the cross-track error due to elongated PSFs in the direction of motion, we focused our analysis on the perpendicular acceleration, and ran our algorithm on all Pan-STARRS tracklets in the ITF regardless of their $\DD$. We found $744$ cases where $\bar{a_{\perp}}$ was $> 3\sigma$ above zero, and extracted and inspected $490$ of them from the Pan-STARRS image archive. The remaining cases were either third party measurements, or were older images not readily available in the archive without special reprocessing.

Of these, $188$ were clearly not real, and as such, we purged them from the ITF database. An additional $140$ tracklets involved two distinct asteroids, and $23$ were poorly measured. Of the remaining, $50$ were fast moving, i.e., trailed detections which would have been previously listed on the NEOCP but were lost due to a lack of follow-up observations, $65$ had no meaningful curvature and were likely selected due to suboptimal astrometry, and $24$ were slow moving but displayed meaningful curvature. The fact that we were able to detect and extract $50$ fast moving objects is promising, as our search did not select based on score or skyplane speed.

Because it can be difficult to find additional confirmation for ITF astrometry, we employed a daily run of our algorithm to search for future objects with curvature $> 2\sigma$. We are hopeful this will result in additional NEO discoveries, but at the very least, will provide an additional dataset of tracklets with meaningful curvature.

The main limitation in this acceleration-based method is that it requires higher-precision astrometry, such as from Pan-STARRS which has $0''.25$ pixels. Tracklets from telescopes utilizing CCDs with larger pixels may not have meaningful measurements of $\bar{a_{\perp}}$. Further, underestimated uncertainties will affect $\bar{a_{\perp}}$, but it is immune to systematic uncertainties.

\section{Discussion}
\label{s:Dicussion}

We have implemented several significant upgrades to the $\D{}$ code. 
These upgrades were much needed and were listed among the key desired developments in \citep{Veres19}. 
Firstly, the update of the population model was imperative due to the continuous increase in the number of discovered objects. Consequently, the undiscovered fraction of objects as a function of $H$ is changing, which impacts the NOID $\DD{}$. This necessitates more frequent updates to the population model than previously executed (in 2015 and 2021).

A second update expanded the list of observatory codes by including additional observatories with their corresponding mean uncertainty errors. This update incorporated newly established observatories that have emerged in recent years, including major survey contributors like ATLAS-South Africa and ATLAS-Chile.

A third update provides the ability for $\D{}$ to parse ADES XML input data, a major direction towards future-proofing the software given the MPC1992 format will eventually become obsolete. As a consequence of this modification, data processing is several times faster than the MPC1992 input. 

Furthermore, the addition of XML ADES-parsing functionality enables the direct ingestion of observer-provided astrometric uncertainties attached to each astrometric position. The availability of astrometric uncertainties has enabled us to analyze the curvature within tracklets. As a result, we introduced a new optional parameter, denoted as $RMS'$, which, if significantly higher than the GCR $RMS$, could indicate the presence of a non-linear in-tracklet motion for a near-Earth object (NEO). This observation suggests that the NEO is in close proximity to Earth, despite having a low-digest score. 

In addition to the implemented $RMS'$, we studied the possibility of finding in-tracklet curvature by fitting the along and cross-track acceleration in a tracklet's motion. Because this method is sensitive to the accuracy of reported astrometric uncertainties, we only studied the acceleration among the Pan-STARRS ITF tracklets for which we have access to their corresponding images and could therefore exclude bad measurements after manual inspection. Our initial findings are promising but require further work and analysis of more data, thus the $\D{}$ implementation is left for a future update. 

As a final addition, we also implemented the ability for $\D{}$ code to read data from \emph{roving observers} (in both MPC1992 and XML format). This addition could be useful for observers that do not have a permanent observatory or belong to the fast growing network of Unistellar telescopes.

Both updates (population + uncertainty configuration file; ADES) were tested on observational data separately. We did not expect a major change in NEO $\D{}$ score behavior but we did anticipate a slight alteration of the undiscovered population to affect the NOID score and for the Raw score to only be affected by the updated uncertainties. A generally smaller variation of $\DD{}-change$ was seen between the XML and non-XML input than the old and new population model and uncertainty configuration file. The results show that the population enhancement could reveal more NEOs but the XML-input would decrease the false-positive rate. Analysis of the ITF initially showed a relatively high number of high-curvature candidates, but closer inspection revealed bad astrometry or reported false detections, because objects in the ITF are not human vetted--unlike the NEOCP candidates.

One of the identified risks in this work is the accuracy of reported uncertainties. That is, while we need to rely on reported values, we have shown that in some cases the values are underreported and in few cases we identified highly exaggerated uncertainties. While the major surveys have switched to ADES-XML reporting almost exclusively, some surveys (such as Purple Mountain, MAP, DECam) still submit in the MPC1992 format. We invite all the submitters to switch to ADES XML and provide as much as infomation possible with their astrometry.  

The upcoming step would be the concurent implementation of the updated $\D{}$ code at the MPC and the NEO community. The $\D{}$ remains a project under ongoing development that could benefit from further enhancements. As such, input from the community on potential improvements and bug identification is highly sought after.

\acknowledgments
This work was supported by the MPC's NASA cooperation agreement funding. We also acknowledge support of the Pan-STARRS project at the University of Hawaii for funding support and support by the Oumuamua-Laukien fellowship awarded to the Galileo Project at Harvard University.
\clearpage 

\clearpage 

\appendix

\section{Software availability}

The $\D{}$ source code is freely available at \url{https://github.com/Smithsonian/digest2}. The software distribution contains the archive of older versions of $\D{}$, the population model with the code to generate the model and the configuration file with the per-obscode uncertainties. 

\bibliography{references}

\section{Example Tracklet - ADES}
\label{app:example_tracklet}
Tracklet of a main-belt asteroid 2022 UG112 in a 80-column format\footnote{\url{https://minorplanetcenter.net/iau/info/ObsFormat.html}}, reported by Mt. Lemmon Survey (G96). It was submitted as an incidental astrometry of an uknown object and later linked with four additional tracklets on three distinct nights. The $\DD{}=0$.

\begin{verbatim}
     C88LQ02 1C2022 10 26.31536003 06 26.176+15 57 49.75         22.33GV     G96
     C88LQ02 1C2022 10 26.32062103 06 25.942+15 57 49.90         21.77GV     G96
     C88LQ02 1C2022 10 26.32588503 06 25.628+15 57 50.58         21.54GV     G96
     C88LQ02 1C2022 10 26.33116703 06 25.348+15 57 50.18         21.63GV     G96
\end{verbatim}

The first detection of the same tracklet in ADES\footnote{\url{https://www.minorplanetcenter.net/iau/info/ADES.html}} format (XML), in addition to previously provided epoch, right ascension, declination, magnitude, band, catalog and observatory code, the XML format also provides information on positional uncertanties (rmsRA, rmsDec, rmsCorr), uncertainty in magnitude (rmsMag), signal-to-noise ratio (logSNR), and conditions plate solution (rmsFit, nStars). Submission format is A17 (ADES 2017).

\begin{lstlisting}[language=XML]
<?xml version='1.0' encoding='UTF-8'?>
<ades version="2017">
  <optical>
    <provID>2022 UG112</provID>
    <trkSub>C88LQ02</trkSub>
    <obsID>LYuC8PF30000EvHV010000Lla</obsID>
    <trkID>00000GurgQ</trkID>
    <mode>CCD</mode>
    <stn>G96</stn>
    <prog>01</prog>
    <obsTime>2022-10-26T07:34:07.063Z</obsTime>
    <ra>46.609065</ra>
    <dec>15.963820</dec>
    <rmsRA>0.267</rmsRA>
    <rmsDec>0.276</rmsDec>
    <rmsCorr>0.0002</rmsCorr>
    <astCat>Gaia2</astCat>
    <mag>22.33</mag>
    <rmsMag>0.304</rmsMag>
    <band>G</band>
    <photCat>Gaia2</photCat>
    <logSNR>0.618</logSNR>
    <rmsFit>0.103</rmsFit>
    <nStars>6616</nStars>
    <subFmt>A17</subFmt>
  </optical>
</ades>
\end{lstlisting}

\section{Examples for a positive curvature detection for low-digest NEOs}
\label{app:example_curvature}
P21CWud  was an ITF precovery tracklet of NEOCP candidate A10SfYx that was designated as a new NEO 2023 BL1. The $RMS=0.30\arcsec$, while $RMS'=0.15\arcsec$.

\begin{verbatim}
     P21CWud  C2023 01 18.37052107 42 00.503+49 48 18.30         21.32wU     F52
     P21CWud  C2023 01 18.38264607 41 59.424+49 48 15.17         21.21wU     F52
     P21CWud  C2023 01 18.39473707 41 58.352+49 48 11.48         21.32wU     F52
     P21CWud  C2023 01 18.40684607 41 57.230+49 48 07.25         21.28wU     F52
\end{verbatim}

The following tracklet was a targeted follow-up of NEO K21U12B, the $RMS=0.53\arcsec$ and the reported uncertainties yield $RMS'=0.20\arcsec$.
\begin{verbatim}
     K21U12B KC2021 11 13.44797004 05 40.04 +20 07 55.0          21.50GV     V06
     K21U12B KC2021 11 13.44161704 05 40.66 +20 07 54.9          21.38GV     V06
     K21U12B KC2021 11 13.44479204 05 40.36 +20 07 54.9          21.73GV     V06
     K21U12B KC2021 11 13.45115004 05 39.87 +20 07 55.2          21.79GV     V06


\end{verbatim}

\bibliography{references}

\end{document}